\documentclass[a4paper,10pt]{article}
\usepackage{amsmath}
\usepackage{color}
\usepackage{amssymb}

\title{A short note on extended probability theory.}
\author{Johan Noldus\footnote{email: johan.noldus@gmail.com}}

\begin{document}
\maketitle
\begin{abstract}
We propose two distinct interpretations of extended probabilities which are realistic for the physical world.
\end{abstract}
\section{Introduction}
Negative probabilities show up in physics when one tries to \emph{covariantly} quantize gauge theories (that is theories for massless particles with integer spin greater than zero); also, they appeared in Wigners quasidistribution when he tried to give a statistical meaning on phase space to the wave function.  Recently, an interpretation for probabilities falling in the range $\left[ -1,1 \right]$ was proposed in \cite{Burgin} but alas this is insufficient, probabilities greater than one should also be dealt with.  The usual way to deal with this in quantum theory is to fix the gauge and define physical states as states which satisfy the appropriate gauge condition.  Those physical states are then shown to define a Hilbert space and as such the problem of negative probabilities doesn't pose itself anymore.  In this note, we propose two interpretations, one is a no nonsense interpretation and just says that all these negative numbers and numbers greater than one are just mathematical expressions and do only indirectly determine a standard probability interpretation which, of course, coincides with the usual one if you start with a standard probability distribution from the beginning.   The second one is inspired by Dirac \cite{Burgin} who compared negative probability with negative money, or a debt or shortage of some kind.  Negative apples do also exist, it just means there is an absence of apples.  This interpretation extends the one in \cite{Burgin} who made the error to compare the opposite of the measurement of a particle with the measurement of an anti particle!  Of course, some of these interpretations allow you to surpass the Bell inequalities (it is well known that extended probability theory allows you to do this, see \cite{Selleri}) given that they involve a measurement inefficiency.  In recent experiments however, one claims to have closed as well the measurement efficiency and the communication loophole which should only leave superdeterminism as a possible candidate.        
\section{The no nonsense interpretation.}
Let $\mathcal{H}$ be some generalized Hilbert space with negative norm states.  We define an operator $A$ on $\mathcal{H}$ to be a physical observable if and only if there exists a left basis\footnote{Let $\mathcal{H}$ be a bimodule over some ring $R$, then $v_{\alpha}$ form a left basis if and only if every element $v$ can uniquely be written as $v = v_{\alpha} r_{\alpha}$ with $r_{\alpha} \in R$.} of normalized eigenvectors $\psi_{\alpha}$ satisfying $\langle \psi_{\alpha} | \psi_{\beta} \rangle = \epsilon_{\alpha} \delta(\alpha - \beta)$, with $\epsilon_{\alpha} = \pm 1$, corresponding to real eigenvalues.  It is very well possible in Clifford quantum mechanics that in finite dimensional spaces a continuum of eigenvalues exist for a Hermitian operator as well as a set of non orthogonal eigenvectors.  For the sequel, it is useful to have a short introduction to Clifford and Quaternion quantum mechanics.
\subsection{Some results from Clifford and Quaternion Quantum Mechanics.}
Real quaternions $Q$ are rather special since the form a division algebra, that is every nonzero element has an inverse.  Denote the quaternion basis over the real numbers by $i,j,k=ij$ satisfying $i^2 = j^2 = k^2=-1$ and $ij + ji = 0$.  Next, define the involution $\tilde{a}$ of a general quaternion $a = a_0 1 + ia_1 + ja_2 + ka_3$ by $\tilde{a} = a_0 1 - ia_1 -ja_2 - k a_3$ then it is easy to verify that $a\tilde{a} = \tilde{a}a = a_0^2 + a_1^2 + a_2^2 + a_3^2$ which is always greater than zero unless $a = 0$.  Take a Quaternion bimodule (a vector space over the Quaternions with left and right multiplication) $\mathcal{H}$, then it is possible to define a scalar product on it satisfying
\begin{eqnarray}
\langle v | v \rangle & \geq & 0 \, \textrm{and equality holds if and only if} \, v = 0 \\*
\langle v | w \rangle & \in & Q \\*
\langle v | w \rangle & = & \widetilde{\langle w | v \rangle} \\*
\langle v | wq \rangle & = & \langle v | w \rangle q \, \textrm{where} \,q \in Q \\*
\langle v | qw \rangle & = & \langle \tilde{q}v | w \rangle.
\end{eqnarray} Let $A$ be an operator on $\mathcal{H}$, then the adjoint $A^{\dag}$ is defined in the usual way.  It is fairly easy to demonstrate that on finite dimensional bimodules any Hermitian operator has a complete set of (left) eigenvectors corresponding to real eigenvalues (just real numbers since $\tilde{a} = a$ implies that $a \in \mathbb{R}$).  A vector $v$ is a left eigenvector of an operator $A$ if and only if there exists a $q \in Q$ such that $Av = vq$.  This result also holds in the infinite dimensional case.  Also, two commuting Hermitian operators are simultaniously diagonalizable.  Quaternion quantum mechanics has no negative probabilities and doesn't really deviate much from standard quantum mechanics.  \\* \\*
The situation is very different when one considers Clifford Quantum mechanics.  Take for example the complex Clifford Algebra generated by the standard $\gamma$ matrices, that is $\gamma^{\mu} \gamma^{\nu} + \gamma^{\nu}\gamma^{\mu} = 2 \eta^{\mu \nu}$ where $\eta^{\mu \nu}$ is a standard Minkowski metric with signature $(+ - - -)$.  The involution we are interested in now, see \cite{Noldus}, is $\widetilde{\gamma^{\mu}} = \gamma^{\mu}$ and $\widetilde{c}$ is just the complex conjugate for $c \in \mathbb{C}$.  A general Clifford element is now made up from $1$, $\gamma^{\mu}$, $\gamma^{[\mu} \gamma^{\nu]}$, $\gamma^{\mu}\gamma^{5}$ and $\gamma^5$ where the latter equals $\gamma^{5} = \gamma^{0}\gamma^{1}\gamma^{2}\gamma^{3}$.  These provide a basis for the complex $4 \times 4$ matrix algebra.
Notice now that $\widetilde{a}a$ is not necessarily equal to $a\widetilde{a}$ and moreover, it is in general not even a real number.  The \emph{scalar part} of $\widetilde{a}a$ is not necessarily a positive number since $(\gamma^{1})^2 = -1$, hence the negative probabilities.  Let now $\mathcal{H}$ be a Clifford bi-module and define a scalar product as before except that now $\langle v | v \rangle$ is an element of the Clifford algebra with a real scalar part which might be vanishing even if $v \neq 0$ (take for example $v = w(\gamma^{0} + \gamma^{1})$ where $\langle w | w \rangle = 1$).  Supposing that the scalar product is nonsigular, meaning that there exists an orthogonal basis $v_i$ satisfying $\langle v_i | v_j \rangle = \delta_{ij}$, then one can define the adjoint $A^{\dag}$ of an operator $A$ in the usual way.  Now, it is fairly easy to see that a Hermitian operator doesn't need to have real eigenvalues, nor does it need to have a basis of left eigenvectors, examples can be found in \cite{Noldus}.  It isn't even necessary that $\tilde{a} = a$ for an eigenvalue $a$!  Therefore, we define an operator $A$ to be \emph{physical} if and only if there exists a left orthonormal basis of eigenvectors corresponding to a set of real eigenvalues (which are obviously unique); obviously physical observables are Hermitian.  Given two commuting physical observables $A$ and $B$, it is \emph{not} necessary that they have a common left eigenbasis corresponding to real eigenvalues.  For example consider an orthonormal left basis $|0 \rangle$ and $|1 \rangle$ and $A |0 \rangle = |0 \rangle \lambda$, $A | 1 \rangle = |1 \rangle \mu$ with $\mu, \lambda \in \mathbb{R}$.  Define the action of $B$ on this basis to be $B |0 \rangle = |0 \rangle \gamma^{0}$ and $B| 1 \rangle = - | 1 \rangle \gamma^{0}$; we now show that $B$ is a physical observable.  Consider the orthonormal left basis given by $|3 \rangle = \frac{1}{2} \left( |0 \rangle (1 + \gamma^{0}) + |1 \rangle (1 - \gamma^{0}) \right)$ and $ |4 \rangle = \frac{1}{2} \left( |0 \rangle (1 - \gamma^{0}) + |1 \rangle (1 + \gamma^{0}) \right)$ then\footnote{The reader easily checks that $|3 \rangle, |4 \rangle$ form a left basis since if $|3 \rangle z + |4 \rangle w = 0$ then $z+w = z - w = 0$ since $\gamma^{0}$ is invertible.  Hence, $z = w =0$; orthonormality follows form $(1 \pm \gamma^{0})(1 \pm \gamma^{0}) = 2 \pm 2\gamma^{0}$ and $(1 - \gamma^{0})(1 + \gamma^{0}) = 0$.} the action of $B$ on this basis is given by $B| 3 \rangle = |3 \rangle$ and $B|4 \rangle = - |4\rangle$ which is what we needed to show.  Notice that $B$ has two eigenbasises corresponding to different pairs of eigenvalues.  Since we agree that only real eigenvalues are to be measured, we call two commuting physical operators \emph{compatible} if and only they have a mutual left orthonormal basis of eigenvectors corresponding to real eigenvalues.  Given that the above counterexample is rather special, we conjecture that generically two commuting physical observables are compatible.
\subsection{Negative probabilities?}     
The dynamics on these ``Hilbert spaces'' is then unitary in a generalized sense but it can be for some state $| \phi \rangle$ that the scalar part of
$$\phi_{\lambda} = \langle \phi | P_{\lambda} | \phi \rangle$$ equals a number greater than one or less than zero where $P_{\lambda}$ is the Hermitian projection operator on the eigenspace of eigenvalue $\lambda$.  For example, if the $| \psi_{\beta} \rangle$ span this eigenspace then
$$P_{\lambda} = \sum \epsilon_{\beta} | \psi_{\beta} \rangle \langle \psi_{\beta}|.$$  The attitude we can take now is that these numbers do not directly define probabilities but probabilities can be derived from any sequence of numbers as follows
$$ P(\phi,\lambda) = \frac{| \phi_{\lambda}|}{\sum_{\mu} |\phi_{\mu} |}$$ provided the sum in the denominator converges and $| \phi_{\lambda}|$ denotes the absolute value of the scalar part of $\phi_{\lambda}$\footnote{In Clifford (but not in quaternion) quantum mechanics, it is possible that $\phi_{\lambda}$ is not a real number.}.  Obviously, this coincides with the standard interpretation if all $\phi_{\alpha} \geq 0$ and therefore we have a standard interpretation for all these numbers outside the range $\left[ 0 , 1 \right]$.  There is nothing mysterious about it, it is plain and simple; you just tell to the dynamics that it is not unitary in a standard way and renormalize\footnote{Note that this interpretation behaves covariantly under generalized unitary transformations since the $\phi_{\lambda}$ remain invariant, hence there is no problem with Lorentz covariance.}.  The reader should remark that the measurement axiom gets a strange twist; suppose you have two \emph{compatible} measurements made at spacelike locations and that one observable $A$ determines Hermitian projection operators $P_j$ and the other $B$ determines projection operators $Q_k$ where the $P_j$ and $Q_k$ are all mutually commuting.  Moreover, let us agree that a measurement of $A$ on a state $| \psi \rangle$ results in a state $P_k | \psi \rangle$ if the corresponding eigenvalue has been measured.  Then, what does one mean with causality or with ``one measurement cannot affect the other one''?  Certainly, if it means that measuring the values of $A$ and $B$ leaves the system in the same state irrespective in which order they have been performed then causality is respected since the projection operators commute.  However, if it would \emph{also} mean that the statistics of the mutual events should be independent of the order in which one measures then our definition would fail in contrast to the standard treatment.  Indeed, in the latter the probability of $P_i$ to be detected before $Q_j$ would be $$ \langle \psi | P_i Q_j P_i | \psi \rangle = \langle \psi | P_i Q_j | \psi \rangle $$ which is clearly symmetric under exchange of $P_i$ and $Q_j$.  In our case however, it would be
$$ \frac{| \langle \psi | P_i Q_j P_i | \psi \rangle |}{\sum_k | \langle \psi | P_i Q_k P_i | \psi \rangle |} \frac{| \langle \psi | P_i | \psi \rangle |}{\sum_k | \langle \psi | P_k | \psi \rangle |} $$ which is not symmetric under the exchange of $P_i$ and $Q_j$.  Of course, when one assumes that $A$ and $B$ will be measured first each with a probability of one half, then this expression becomes symmetric again (but here we don't need the commutativity of both operators)\footnote{The local statistics for $A$ remain the same since summing the previous expression over all $Q_j$ leaves just $\frac{| \langle \psi | P_i | \psi \rangle |}{\sum_j | \langle \psi | P_j | \psi \rangle |}$ but the local statistics for $B$ doesn't!  If one were to take the ``same'' observables (we assume a time translation symmetry here) over many pairs of spatially separated regions, and repeat the same experiment, then this could be detected.  But this is not a violation of local causality of course.}.    Note however that for product states $| \psi \rangle \otimes |\phi \rangle$ the individual statistics remain invariant and the joint statistics does not depend upon the order in which has been measured since the previous formula reduces to
$$ \frac{|\langle \psi | Q_j | \psi \rangle|}{\sum_k |\langle \psi | Q_k | \psi \rangle|} \frac{|\langle \phi | P_j | \phi \rangle|}{\sum_k |\langle \phi | P_k | \phi \rangle|}$$ so in ordinary life we don't see such effects\footnote{The local commutativity statement in quantum field theory seems more important for sake of having a Lorentz invariant scattering matrix (as Weinberg seems to suggest \cite{Weinberg}) than as a statement of compatible observables since a local field is not a realistic observable at all (it doesn't even have eigenstates with a definite particle number)!  Indeed, realistic experiments satisfy more stringent criteria than compatibility; it is moreover assumed that there exists a basis of one particle states such that if a state is measured by one apparatus, it cannot be mesured by the other one (this is the axiom of seperable systems).  Field theory is more like a hidden variables programme which merely serves to facilitate computations which do not explicitely depend upon them (such as is the case for the $S$-matrix which only depends upon the Hamiltonian).}.  We therefore assume that causality simply means our first statement that the outcome of both experiments does not depend upon the order in which they were performed (if this were not true, then it would be possible to send a signal between both events).  Note however, that the inertial frame at hand (which determines how the collapse of the wavefunction takes place) has an influence on the probabilities of joint events (if we would repeat a Bell experiment a sufficient number of times).  However, this is natural from the point of view of semiclassical quantum gravity since a measurement is also going to change the evolution of the gravitational field which is in principle something which can be measured, see \cite{Noldus} for more on this.  Obviously, this interpretation does not allow one to surpass the Bell or CHSH inequalities.    
\section{A more subtle interpretation.}
This interpretation is more subtle and therefore open to more scrutiny; we want to take Dirac's idea about negative money to the real physical world.  One therefore reinstates Dirac's hole interpretation (and generalizes it to bosons) but this time not referring to the hole as the anti particle.  That is, we launch the idea that for every pair consisting of an elementary particle and detector, there exists a color $n \in \mathbb{Z}$ indicating the detector's response to that particle.  Suppose you have a detector who has a ground state zero and where state $n > 0$ indicates absorbing $n$ particles and $-n$ the \emph{absence} of $n$ particles relative to the detector's ground state.  An event is simply such a measurement and therefore one particle can have an infinite number of detector responses; for example, an electron arriving at a screen might by absorbed by the screen and release another few electrons from their atomic bounds while those empty places are filled in with electrons from the environment (measurement $+n$) or it might reflect and knock out a few other electrons (measurement $-n$).  Notice therefore that a single particle state might not in some circumstances be distinghuishable from a multi particle state experimentally which is well known since detectors give classical signals and do not directly adress the number of particles involved.  Actually, standard quantum field theory already allows for this by looking for observables which are not diagonal in the particle basis (but here a two particle state is \emph{possibly} distinghuished\footnote{This doesn't need to be so, it might very well be that the statistics is the same albeit the wave functions of the two particles are different.} from a single particle state by means of statistics of detector responses).  Then, the probability for an electron to be absorbed by the screen might be defined as 
$$P(a) = \sum_{\textrm{events} \, \alpha} \frac{1}{N} \textrm{color}_{\alpha}$$ where $N$ is the number of independent single particles you shoot at the screen.  Realistically, if you could distinguish those states, you interpret the state of affairs of course in terms of positive probabilities
$$P(n) = \frac{N_n}{N} $$ where $N_n$ equals the number of measurements of state $n$ for $n \in \mathbb{Z}$ which means the theory would have lost predictivity which was also the case in the interpretation of \cite{Burgin} (you may of course suggest that $\lim_{N \rightarrow \infty} P(n)$ does not exist but that would be very nontrivial from a physicist's point of view).  Suppose now that \emph{you} as observer would \emph{not} be able to distinguish those states and consider them all equal except for zero which means the detector remained in its ground state then all you could see is $P(0)$ or equivalently $1 - P(0)$, which is then interpreted as the probability of absorption.  In that case the negative probability theory just gives as much information as your observations do, albeit based upon more detailed \emph{assumptions} (of course, the negative probability theory does not contain those elements about the measurement apparatus).  You might even be willing to refute the theory based on this mismatch between $1-P(0)$ and $P(a)$ or you might want to construct more sensitive detectors in which case you might want to conclude that you need a more predictive theory (if it exists).  Such loss of information was to be expected since having $1000$ euro to buy stuff with tomorrow is the same as having $1200$ euro now with an instantaneous debt of $200$ euro to be paid off today.  This is not a very satisfactory state of affairs but perhaps it works like that; actually such information loss occurs all the time when we make interpretations about the physical world.  For example, this might be so even for the double slit experiment where we count black dots on a white background; it is by no means certain that one black dot corresponds to the absorption of a single particle there, it might be multiple particles or the extensiveness of the dot might be due to a particular backreaction of the detector.  Also, it is possible that when measurement occurs the detector can be in multiple states corresponding to a single particle state which cannot be described by a particle observable since it is oblivious to the detector state.  Here one should consider the composite system, particle and detector together. \\* \\* In the philosophy of this section, a theory with complete knowledge about the state of affairs should have positive probabilities only, but quantum mechanics isn't such a theory and therefore negative probabilities might find a suitable place in it.
\section{Conclusions.}
In this note, we proposed two interpretations for the numbers which show up in computations and which are expected to have an immediate probability interpretation.  One attitude is that they dont have a direct probability interpretation in the general case and a probability distribution needs to be defined from them.  This interpretation is conceptually clear and refers directly to measurable quantities.  The second interpretation however assumes that your theory can only predict limited information about the system at hand and is by no means complete unless there is a principle of nature forbidding us to do any better.  This approach draws attention to the fact that we have to be careful when dealing with determining particle properties from responses of macroscopic apparati, there are always assumptions involved using our knowledge about such systems, that is how our senses percieve them. \\* \\*  What this work suggests is that it is by no means \emph{necessary} to have a gauge symmetry which turns all these numbers into positive numbers again unless some other sacred principle of nature requires it.  This is even not the case in \emph{field theory} for massless spin one and spin two particles since positive probabilities are here assumed from the outset (or else gauge invariance is assumed).  In \cite{Noldus}, we have a locally Lorentz covariant theory with negative ``probabilities'',  but all particles come from standard (positive probability) Hilbert space representations of the (local) Poincar\'e group, so there is no such need to deal with them in this case.  It is just the number system which causes these negative ``probabilities'' to arise and this should not have any serious impact on the theory.    


\begin{thebibliography}{99}
\bibitem{Burgin} M. Burgin, Interpretations of Negative probabilities,  arXiv:1008.1287. 
\bibitem{Noldus} J. Noldus, Foundations of a theory of quantum gravity, arXiv:1101.5113.
\bibitem{Selleri} A. Afriat and F. Selleri, The EPR paradox in Atomic, Nuclear, and Particle Physics, Plenum Press, New York and London.
\bibitem{Weinberg}  S. Weinberg, The quantum theory of fields, foundations, Cambridge University Press.
\end{thebibliography}
\end{document}